\documentstyle[12pt,a4wide]{article}
\begin{document}
\vspace{24pt}
\begin{center}
{ \Huge
{\bf General D-branes Solutions From}}
\vspace{12pt}
{ \Huge{\bf String Theory}}
\end{center}
\vspace{88pt}
\begin{center}
\underline{ABSTRACT}
\end{center}

We give the general solution for the elementary and solitonic D-brane
configurations as a result of a reinterpretation of the already known
p-branes. These solutions are found by means of a relevant conformal
transformation on the string inspired action and its dual form. From this
point of view, the nature of the electric and magnetic charge is clearer
and the elementary and solitonic behaviour dependence on the initial
lagrangian set. We give a complete characterisation of the spacetime
defined by these solutions. The dual pair of instanton and 7-brane
solution is presented as an example. 

\pagebreak

\newpage
\section{Introduction}

Recent developments on string theory and duality conjectures suggest the
existence of an underlying theory, so-called M-theory, such that the
different string theories are just perturbative approximations of it. From
the standpoint of a given string theory, the other formulations are seen
as solitonic non-perturbative states (for a review see \cite{sch1}). Apart
from these 1-dimensional extended objects there are other p-brane
configurations which have a very important role in the understanding of
the new picture of M-theory. These solitonic states are especially
interesting as they conserve half of the initial supersymmetry, where the
remaining non-broken symmetries are related to the existence of a
$\kappa$-symmetry, a generalisation of the well known Green-Schwartz
world-sheet $\kappa$-symmetry. 

In Type II theories, the massless bosonic states include the so-called
Ramond-Ramond (RxR) sector on the top of the more conventional
Neveu-Schwarz-Neveu-Schwarz (NSxNS) sector, graviton and dilaton. 

Recently, the p-branes carrying the Ramond-Ramond sector have been
associated with superstring configuration known as Dirichlet branes
\cite{p1}. From this point of view, there should be solitons for all the
values of p from -1 to 9. This range is related to the expansion of the
Ramond-Ramond bispinor in string theory as a Lorentz tensor. The allowed
values for the dimension of the relevant RxR form are 0, 2, 4 for Type IIA
theory and 1, 3, 5 for Type IIB theory. This formulation also admits the
so-called dual RxR forms. 
 
The following solutions are already known: the 0-brane or black hole, the
4-brane ,the 5-brane and the 6-brane solutions of Type IIA, the 1-brane o
string , the self-dual 3-brane, 5-brane , the (-1)-brane or instanton and
its dual the 7-brane \cite{p2,p3,to,b,e,be2,gi} . It is worth mentioning
that in fact there are an infinite set of SL(2,Z) multiple of both
``dyonic'' 1-branes \cite{sch2} and 5-branes in the Type IIB theory (for a
review see \cite{du}). 

In this paper we give an explicit form for the general solution of all the
D-branes in Type IIA and Type IIB theories. Although the construction
resembles very much the previous general solutions for p-branes \cite{ke},
the D-p-branes have fascinating new features, like the non-singularity of
some solutions in the relevant frame (either string or Einstein frame),
real wormholes and connections with the cosmic string, to mention a few. 

Some of the general forms for these solutions can be found in previous
works \cite{to,be2,be1}, where the focus was more into either the relation
between supermembrane theory and superstring Type IIA theory or the
T-duality between the different D-p-branes solutions. Here the emphasis is
given to a reinterpretation from the point of view of p-branes as
solutions of effective lagrangians with RxR forms and its dual
formulations. In particular, this framework shows the dependence of the
elementary or solitonic characteristic on the initial lagrangian
\cite{du}. 

In section 2 we present the lagrangian corresponding to Type II theories.
Then, by means of a reinterpretation and a conformal transformation
inspired on the string nature of the lagrangian, we recover the general
form for the lagrangians for p-branes. In section 3 we give a short review
of the general solution for the p-branes, the dual relations between the
elementary and solitonic solutions on the same theory and the relation
between solutions on dual theories. In section 4 we give the general
properties for the solitonic solutions placing special emphasis on the
solution with d=8 and its dual formulation.

\section{The General D-p-Brane Solution}

The effective lagrangian for the massless modes in Type II theories is
given by the expression,

\begin{equation} S_{10}=\int d^Dx\sqrt{-g}(e^{-2\phi}(R
+4(\partial\phi)^2)) - \frac{1}{2(d+1)!}F_{d+1}^2) \end{equation} where
the NSxNS field is not considered and the values of $d$ depend on which
version of the Type II theory has been used , basically we have
$d+1=0,2,4$ for Type IIA and $d+1=1,3,5$ for Type IIB. 

We are interested in finding out solutions that could be interpreted as
D-branes solutions. The key idea is to generalise the above lagrangian to
get a new formulation that will give us a better known theory. Therefore,
we consider a conformal transformation which takes us to the so called
Einstein frame. 

\begin{equation} S_{10}=\int d^Dx\sqrt{-g}(e^{-2\phi}(R
+4(\partial\phi)^2)) - \frac{1}{2(d+1)!}e^{a(d)'\phi}F_{d+1}^2)
\end{equation} We note that $a(d)'$ depends on the dimension of the
background space $D$ and on the dimension of the field strength.
Explicitly we have,

\begin{equation} a'=\frac{D}{4}-\frac{d+1}{2} \end{equation} To obtain the
well known lagrangian for the p-branes, $a(d)'$ should be equal to the
characteristic coupling constant $a(d)$, defined in terms of $d$ and
$\bar{d}$ ( dual dimension ) by the expression,

\begin{equation} a=-\sqrt{4-\frac{2d\bar{d}}{d+\bar{d}}} \end{equation} As
a result, $a(d)'$ must satisfy two different algebraic equations: the
equation which takes us to p-brane lagrangians and the restriction that
takes us to Type II theory. These two algebraic equations can be written
simultaneously by means of the equation,

\begin{equation} |a(d)|=\frac{1}{2}(4-d) \end{equation} Hence, we have the
allowed range $d\geq4$, for the corresponding solutions of the Type II
theory. This restriction in the possible values of $d$ can be overcome by
considering as starting point a positive value for $a(d)$, so that the
parameter $a(d)'$ goes to $-a(d)'$, recovering the sector $d\leq4$. We
will see that this change in $a(d)$ corresponds to the dual formulation of
the Lagrangian for the p-branes. Therefore the character of the solutions
will be, of course, reversed in the sense that the solitonic and
elementary solutions will exchange places. 

\section{The General p-brane Solution}

From the Heterotic string effective action for the bosonic sector, we can
generalise to \cite{ke}
 
\begin{equation} S_{D}=\frac{1}{2k^2}\int d^Dx\sqrt{-g}( R
-\frac{1}{2}(\partial\phi)^2) - \frac{1}{2(d+1)!}e^{a(d)\phi}F_{d+1}^2)
\end{equation} where we consider a spacetime background of dimension D, in
which we have a (p+2)-form field strength $F_{p+2}$ interacting with
gravity , $g_{MN}$ and a dilaton field $\phi$, via the above action.
$a(d)$ is a yet undetermined negative constant. 

The dual dimension $\bar{d}$ is defined as $\bar{d}=D-2-d$ and the
magnetic and electric charges as

\begin{equation} e_{d} = \int_{S^{\bar{d}+1}} e^{-a(d)\phi} F^{*}
\;\;\;;\;\;  g_{\bar{d}} = \int_{S^{d+1}} F \end{equation}

To solve the above equations we consider the most general split invariant
under $P_{d}xSO(D-d)$, where $P_{d}$ is the d-dimensional Poincar\'{e}
group. We split the coordinates as $x^{M}=(x^{\mu},x^{m})$ where
$\mu=0,1...p=(d-1)$ and $m=d,...D-1)$. For the dilaton we consider
dependence solely on the transverse coordinates, $\phi=\phi(r) $ where
$r=\sqrt{y^{m}y^{m}}$ and the line element is given by,

\begin{equation}
ds^{2}=e^{2A(r)}dx^{\mu}dx_{\mu}+e^{2B(r)}dy^{m}dy_{m}
\end{equation}
For the strength we define,

\begin{equation}
F_{d+1}=dA_{d}\;\;;\;\;\;A_{\mu_{1}.......\mu_{d}}=\epsilon_{\mu_{1}.......\mu{d}}e^{C(r)}
\end{equation} For the elementary solution, or

\begin{equation}
F_{\mu_{1}.......\mu_{d+1}}=\lambda\epsilon_{\mu_{1}.......\mu_{d+1}l}\frac{y^{l}}{r^{d+2}}
\end{equation} for the magnetic solution. 

After imposing the necessary conditions to ensure supersymmetry
conservation, we find that the general solution for the elementary case is

\begin{eqnarray} &&ds^{2}=H^{\frac{-\bar{d}}{d+\bar{d}}}dx^{\mu}dx_{\mu}+
H^{\frac{d}{d+\bar{d}}}dy^{m}dy_{m} \\ &&e^{\phi}=H^{\frac{ a}{2}} \\
&&H(r)=\left\{ \begin{array}{ll}
	1+\frac{K_{d}}{r^{\bar{d}}}\;\;\;  \mbox{if $\bar{d}>0$}\\
	C_{o}+K_{o}ln(r)\;\;\;\mbox{if $\bar{d}=0$}
	\end{array}
	\right. 
\end{eqnarray}

where $a=-\sqrt{4-\frac{2d\bar{d}}{d+\bar{d}}}$ 
                
To get the solitonic solutions, one replaces $d$ by $\bar{d}$ and set
$a(\bar{d})=-a(d)$. This will give a $\bar{d}$-brane magnetically charged.
In the above solutions, the relation between electric and magnetic charge
is given by $ e_{d}g_{\bar{d}}=2\pi n$, for n integer. 

The p-branes solutions presented before were regarded as solutions for the
NSxNS sector. It is well known that these corresponding lagrangians can be
rewritten in terms of dual objects where the relevant conformal factor
must be taken into consideration. As a general rule we have,

\begin{eqnarray}
&&\bar{F}_{\bar{d}+1}=e^{a(d)\phi}F_{d+1} \\
&&g^{d}_{MN}=e^{-a(d)\phi/d}g^{E}_{MN}  \\
&&g^{\bar{d}}_{MN}=e^{a(d)\phi/\bar{d}}g^{E}_{MN} 
\end{eqnarray}

where $g^{E}_{MN}$ stands for the Einstein metric.  Hence, the above
conformal factors are the ones to be used if we want to consider change of
frames from the Einstein picture to the d-brane picture. 

From the analysis of section 2 and the specific form for the general
p-branes, we get explicity both types of D-p-brane solutions, elementary
and solitonic. 

\vspace{12pt}
For the elementary case, we have

\begin{eqnarray}
&&ds^{2}=H^{-1/2}dx^{\mu}dx_{\mu}+ H^{1/2}dy^{m}dy_{m} \\
&&e^{\phi}=H^{1-d/4} \\
&&H(r)=\left\{ \begin{array}{ll}
	1+\frac{K_{d}}{r^{\bar{d}}}\;\;\;  \mbox{if $\bar{d}>0$}\\
	C_{o}+K_{o}ln(r)\;\;\;\mbox{if $\bar{d}=0$}
	\end{array}
	\right. \\
&&F_{\mu_{1}.......\mu_{d}m}=\epsilon_{\mu_{1}.......\mu_{d}}\partial_{m}H^{-1}
\end{eqnarray}
where $\mu=0,...,d-1$, $\;\;m=d,...,9$.

\vspace{12pt} 
For the the solitonic case

\begin{eqnarray}
&&ds^{2}=H^{-1/2}dx^{\mu}dx_{\mu}+ H^{1/2}dy^{m}dy_{m} \\
&&e^{\phi}=H^{d/4-1} \\
&&H(r)=\left\{ \begin{array}{ll}
	1+\frac{K_{\bar{d}}}{r^{d}}\;\;\;  \mbox{if $d>0$}\\
	C_{o}+K_{o}ln(r)\;\;\;\mbox{if $d=0$}
	\end{array}
	\right. \\
&&F_{m_{1}.......m_{d+1}}=\lambda\epsilon_{m_{1}.......m_{d+1}l}\frac{y^{l}}{r^{d+2}}
\end{eqnarray} where $\mu=0,...\bar{d}-1$, $\;\;m=\bar{d},...,9$. 

\section{General behaviour and Singularities Issues}

To consider the issue of singularity we must recalled that there are
basically three different types of frames relevant in the general
formulation of the theory: the Einstein frame, the string frame and the so
called source frame or sigma-model frame. The last frame happens to be
very important as the whole picture of duality between different p-branes
actions becomes clearer \cite{du}. Then, we can determine the relevant
geometric invariant like the Ricci scalar on all three different frames.
Obviously, on the sigma-model frame is were, we should classify a given
solution as elementary or solitonic. In our case the relation between the
frame where the Type II solutions are given ( string frame) and the
sigma-model frame is,

\begin{eqnarray} &&g^{d}_{MN}=e^{(\frac{1}{2}+a(d)/d)\phi}g^{\sigma}_{MN}
\\
&&g^{\bar{d}}_{MN}=e^{(\frac{1}{2}+a(\bar{d})/\bar{d})\phi/}g^{\sigma}_{MN}
\end{eqnarray} If we compute the curvature invariant for the solitonic
case in string frame, some of the solutions will appear singular,
depending on the dimension of the brane under consideration. On the other
hand, if we calculate the curvature invariant on the sigma-model frame, no
singularities are found as expected. The different results are show below. 

\vspace{12pt} \begin{center} \begin{tabular}{|c|c|c|c|} \hline brane
dimension $d$&Einstein frame&string frame&sigma--model frame\\ \hline
\hline $d<4$&$R\rightarrow\infty$&$R\rightarrow\infty$&$R\rightarrow
const.$\\ \hline $d=4$&$R\rightarrow const.$&$R\rightarrow
const.$&$R\rightarrow const.$\\ \hline
$d>4$&$R\rightarrow\infty$&$R\rightarrow 0$&$R\rightarrow const.$\\ \hline
\end{tabular} \end{center} \vspace{12pt} The family of metrics presented
here brings pathological behaviours at the limit $r\rightarrow0$. To
obtain the nature of the hypersurface $r=0$, different change of
coordinates should be done. The analysis on the sigma-model frame
\cite{du2}, gives that the metric goes to $(AdS)_{d+1}xS^{10-d}$, as
$r\rightarrow0$, and goes to flat Lorentzian space as $r\rightarrow\infty$
interpolating two vacuum solutions. Also the hypersurface $r=0$ is a
degenerated event horizon, where a continuation through $r<0$ can be done.
The new region happens to be isomorphic to the original region. This
behaviour was expected as it can be shown that these solutions resemble
the extreme cases already discovered some time ago \cite{gi2}. We should
mention that the dilaton is singular at the horizon for $d>4$, hence
srictyly speaking non of these solutions are solitonic, nevertheless in a
few cases the singularity can be overcome by interpreting these solutions
as a dimensional reduction from 11D supergravity \cite{du2}. 

The above analysis was based on general characteristics of the solutions,
in particular we can consider some features which depend on the dimension
of the brane. First of all, the case $d=4$ is selfdual and corresponds to
the solution of the normal p-branes, since the conformal factor gives the
identity. From the elementary branch, the case $d=0$ is very interesting,
for we are in presence of the instanton found by Gibbons et. all
\cite{gi}, which is a wormhole. From the solitonic branch, we pay
attention to the $d=8$ ($\bar{d}=0$). This is another wormhole solution
such that the conserved charge is a topological charge. This wormhole
corresponds to the previous wormhole but, in the dual theory. Both
solutions are given by the equations

\begin{eqnarray}
&&ds^{2}=H^{1/2}dy^{m}dy_{m} \\
&&e^{\phi}=H \\
&&H(r)=1+\frac{K_{0}}{r^{8}}\\ 
&&F_{m}=\partial_{m}H^{-1}
\end{eqnarray}
where $\;\;m=0,...,9$.   and 

\begin{eqnarray} &&ds^{2}=H^{1/2}dy^{m}dy_{m} \\ &&e^{\phi}=H^{-1} \\
&&H(y)=1+\frac{K_{\bar{0}}}{y^{8}} \\
&&F_{m_{1}.......m_{8}}=\lambda\epsilon_{m_{1}.......m_{8}l}\frac{y^{l}}{r^{10}}
\end{eqnarray} where $\;\;m=0,...,9$. 

Note that in the Einstein frame the elementary solution gives flat space
while the solitonic solution does not. On the top of these solutions we
found their dual configurations, the elementary and solitonic 7-branes. 

The elementary 7-brane is given by,

\begin{eqnarray}
&&ds^{2}=H^{-1/2}dx^{\mu}dx_{\mu}+ H^{1/2}dy^{m}dy_{m} \\
&&e^{\phi}=H^{-1} \\
&&H(r)=C_{o}+K_{o}ln(r)\\ 
&&F_{m}=\partial_{m}H^{-1}
\end{eqnarray}
where $\mu=0,...7$, $\;\;m=8,9$.

In the Einstein frame the line element gives,

\begin{equation}
ds^{2}=dx^{\mu}dx_{\mu}+ H^{1/2}dy^{m}dy_{m} 
\end{equation}

while the solitonic solution is given by,

\begin{eqnarray} &&ds^{2}=H^{-1/2}dx^{\mu}dx_{\mu}+ H^{1/2}dy^{m}dy_{m} \\
&&e^{\phi}=H \\ &&H(r)=C_{o}+K_{o}ln(r)\\
&&F_{m_{1}.......m_{8}}=\lambda\epsilon_{m_{1}.......m_{8}l}\frac{y^{l}}{r^{10}}
\end{eqnarray} where $\;\;m=0,...,9$. 

Finally in the Einstein frame the line element is

\begin{equation}
ds^{2}=H^{-1}dx^{\mu}dx_{\mu}+ dy^{m}dy_{m} 
\end{equation}

\section{Conclusions}

We have seen how the general form for the solutions to Type II theory can
be obtained from the already known conventional solutions of p-branes. The
method is based on a reinterpretation of the conformal transformations
which links dual formulations within the p-brane theories. This
reinterpretation gives rise to a conformal transformation related to a
unique string frame, which, after an algebraic imposition on the factor
$a(d)$, reproduces the solutions for Type II theory. The fact that this
constraint in $a(d)$ is compatible with its form from the p-brane
solutions is what guarantees the consistency of the whole procedure. 

The solutions presented here happen to be elementary or solitonic for new
values of $d$. This behaviour is explained by the fact that these
solutions come from different initial theories which are duals. This point
of view follows from discussions in \cite{du}. 

As an example, we have recovered explicitly some of the already known
solutions (instanton and 7-brane) and we have also obtained their dual
formulations as a gift from the generality of the method. All the possible
solutions are contained here, together with the corresponding duals. 

As a last remark, although the majority of these solutions were found
already in the literature, the novelty of these method brings together
mathematical structure of p-branes and D-branes. 

\vspace{24pt}
{\bf Acknowledgements}
\vspace{24pt}

We thanks Dr I. Moss for calling my atention on this subject and for
useful discussions. This work was suported by the Venezuelan Commision on
sciencies CONICIT. 
 
\pagebreak

\end {document}